\documentclass{PoS}

\def\fig#1{fig.~{\ref{#1}}}
\def\Fig#1{Fig.~{\ref{#1}}}
\def\figs#1#2{figs.~{\ref{#1}} and {\ref{#2}}}

\def\pb{\bar p}

\def\ETslash{{\s E}_T}

\def\Wjj{$W\,\!+\,2$}
\def\Wjjj{$W\,\!+\,3$}

\def\Wmjj{$W^-\,\!+\,2$}
\def\Wmjjj{$W^-\,\!+\,3$}

\def\Zj{$Z\,\!+\,1$}
\def\Zjj{$Z\,\!+\,2$}
\def\Zjjj{$Z\,\!+\,3$}

\def\Zjjja{$Z\,\!+\,1,2,3$}

\def\Vjjj{$V\,\!+\,3$}
\def\Vjn{$V\,\!+\,n$}
\def\Vjnp#1{$V\,\!+\,(n+#1)$}
\def\Vjjx{$V\,\!+\,2,3$}

\def\jet{{\rm jet}}

\def\NLO{{\rm NLO}}
\def\tree{{\rm tree}}
\def\oneloop{{1\hbox{-}{\rm loop}}}
\def\Obs{{\rm Obs}}

\def\HTp{\hat H_T}

\def\BlackHat{{\sc BlackHat}}
\def\BlackHatT{B{\Large LACK}H{\Large AT}}
\newbox\charbox
\newbox\slabox
\def\s#1{{      % Feynman slash
        \setbox\charbox=\hbox{$#1$}
        \setbox\slabox=\hbox{$/$}
        \dimen\charbox=\ht\slabox
        \advance\dimen\charbox by -\dp\slabox
        \advance\dimen\charbox by -\ht\charbox
        \advance\dimen\charbox by \dp\charbox
        \divide\dimen\charbox by 2
        \raise-\dimen\charbox\hbox to \wd\charbox{\hss/\hss}
        \llap{$#1$}
}}

\def\nuornub{{}^{\raise1.3pt\hbox{\ttiny(}}\hskip -0.2pt\overline{\kern 
-0.5pt \nu \kern -0.4pt}\hskip 0.3pt{}^{\raise1.3pt\hbox{\ttiny)}}}

\title{
\rule{0cm}{2.5cm}\vspace{-3.5cm}\\{ 
\it \normalsize IPPP/09/96
\hfill MIT-CTP-4095  \hfill Saclay--IPhT--T09/268  \\
SLAC--PUB--13868 \hfill SB/F/371-09 \hfill UCLA/TEP/09/202 
 } 
\\ \vspace{2cm}
Next-to-Leading Order Jet Physics with \BlackHatT}

\ShortTitle{NLO Jet Physics with \BlackHat \hskip 7.4 cm 
Kosower and Ma\^{\i}tre}

\author{C.~F.~Berger${}^{a}$, Z.~Bern${}^b$,
L.~J.~Dixon${}^c$, F.~Febres Cordero${}^d$, D.~Forde${}^{e,f}$,
T.~Gleisberg${}^c$, \hskip .3 cm  H. Ita${}^b$,
D.~A.~Kosower${}^{g*}$\,
and D.~Ma\^{\i}tre${}^{h} \! \! \!$ 
\speaker{}
\\
${}^a$Center for Theoretical
Physics, MIT,
      Cambridge, MA 02139, USA \\
${}^b$Department of Physics and Astronomy, UCLA, Los Angeles, CA
90095-1547, USA \\
${}^c$SLAC National Accelerator Laboratory, Stanford University,
             Stanford, CA 94309, USA \\
${}^d$Universidad Sim\'on Bol\'{\i}var, Departamento de
F\'{\i}sica, Apartado 89000, Caracas 1080A, Venezuela\\
${}^e$Theory Division, Physics Department, CERN, CH--1211 Geneva 23, 
    Switzerland\\
${}^f$NIKHEF Theory Group, Science Park 105, NL--1098~XG
  Amsterdam, The Netherlands\\
${}^g$Institut de Physique Th\'eorique, CEA--Saclay,
          F--91191 Gif-sur-Yvette cedex, France\\
${}^h$Department of Physics, University of Durham,
          DH1 3LE, UK
}

\abstract{
We present several results obtained using the \BlackHat{}
next-to-leading order QCD program library, in conjunction with SHERPA.
In particular, we present distributions for vector boson 
plus 1,2,3-jet production at the Tevatron and at the asymptotic running
energy of the Large Hadron Collider, including new \Zjjj-jet distributions.
The \Zjj-jet predictions for the second-jet $P_T$
distribution are compared to CDF data.  We present the jet-emission
probability at NLO in \Wjj-jet events at the LHC, where the tagging jets are
taken to be the ones furthest apart in pseudorapidity.
We analyze further the large left-handed $W^\pm$ polarization,
identified in our previous study, for $W$ bosons produced at high
$P_T$ at the LHC.
}

\FullConference{RADCOR 2009 - 9th International Symposium on 
                Radiative Corrections
                (Applications of Quantum Field Theory to Phenomenology),
		 October 25 - 30 2009, 
		 Ascona, Switzerland}

\begin{document}

\section{Introduction}

The dawn of the Large Hadron Collider (LHC) era
brings renewed incentive to continue improving theoretical predictions
of Standard-Model backgrounds to new physics searches.  For many 
searches, including some channels for the Higgs boson and for dark matter
particles, the signals will be excesses in jet\,+\,lepton 
or jet\,+\,missing~$E_T$ distributions.  Such signals can be mimicked
by Standard-Model processes; accordingly, a thorough and quantitatively
reliable theoretical prediction is needed.  
This requires a calculation through next-to-leading order (NLO) in QCD.

Leading-order (LO) computations, while an important first step,
suffer from a strong dependence on the unphysical renormalization and
factorization scales.  At this order, they enter only through the
strong coupling $\alpha_s$ and parton distribution functions,
uncompensated by any behavior of the short-distance partonic matrix
elements.  Because the QCD coupling is large and runs quickly, the
absolute normalization of cross sections has a substantial dependence on
scales.  For reasonable scale variations, the dependence is
of the order of $\pm40\%$ for the \Vjjj-jet processes we shall study,
with $V$ a heavy electroweak vector boson.  The dependence also grows
substantially with increasing number of jets.  At NLO, the virtual
corrections introduce a compensating dependence on the scales. The
scale dependence shrinks to $\pm10\%$, and we obtain a quantitatively
reliable answer.
Shapes of distributions can also show a dramatic scale dependence with poor
scale choices. Some shapes do display noticeable ``genuine''
NLO corrections, independent of scale issues.

NLO predictions for \Vjn-jet production at hadron
colliders require several ingredients:
\begin{itemize}
\item tree-level \Vjnp{2}-parton matrix elements, which provide
the LO contribution;

\item interference of one-loop and tree amplitudes for
\Vjnp{2} partons (virtual contribution);

\item tree-level \Vjnp{3}-parton matrix elements
(real-emission contribution);

\item a subtraction approximation capturing the singular behavior of
the real-emission term;

\item the integral of the approximation over the singular
phase space (real-subtraction term).
\end{itemize}
These contributions must be convoluted with
parton distribution functions, obtained from NLO fits, and
integrated over the final phase space, incorporating appropriate
experimental cuts.

Schematically, we combine the contributions as follows,
\begin{eqnarray}
\frac{d\sigma_{V+n}^\NLO}{d\Obs}
= \int dx_{1,2} f_{1}f_{2} &&
   \biggl[\int d\Phi_n\,\delta_{\Obs}\,\sigma^\tree_{2\rightarrow V+n}
 +\int d\Phi_n\,\delta_{\Obs}\,\bigl(\sigma^\oneloop_{2\rightarrow V+n}
                               +\sigma^{\int\rm app}_{2\rightarrow V+n}\bigr)
\nonumber\\&&
 +\int d\Phi_{n+1}\,\delta _{\Obs}\,
    \bigl(\sigma^\tree_{2\rightarrow V+n+1}
          -\sigma^{\rm app}_{2\rightarrow V+n+1}\bigr)
   \biggr] \,,
\label{BasicNLODistribution}
\end{eqnarray}
where $d\Phi_n$ denotes the $V+n$-parton phase space; $dx_{1,2} f_1
f_2$ the integral over the appropriate parton distributions, a sum
over types being implicit; $\delta_{\Obs}$, the binning function for
the desired distribution;
$\sigma^\tree$, the tree-level squared matrix elements;
$\sigma^\oneloop$, the virtual corrections; $\sigma^{\rm app}$, the
approximation to the real-emission contribution; and $\sigma^{\int\rm
app}$, the approximation's integral over singular phase space.  The
set of subtraction terms ensures 
that each of the
terms in this equation is separately finite, and thus may be computed
numerically.

We use the \BlackHat{} program
library~\cite{BlackHatI, ICHEPBH, PRLW3BH, W3jDistributions} 
to compute the virtual
corrections $\sigma^\oneloop$, and the SHERPA package~\cite{SHERPA} to
compute $\sigma^\tree$ and the
required approximation~($\sigma^{\rm app}$ and $\sigma^{\int\rm app}$). 
The approximation uses the Catani--Seymour dipole
approach~\cite{CS}.  The phase-space integration is 
performed with SHERPA, implementing a multi-channel
approach~\cite{MultiChannel}.

% What BlackHat does
The \BlackHat{} library implements on-shell methods
for one-loop amplitudes numerically.  Such amplitudes can be written
as a sum of cut terms $C_n$, containing branch cuts in kinematic
invariants, and rational terms $R_n$, free of branch cuts,
\begin{equation}
A_n\ =\ C_n\ +\ R_n\,.
\label{CutRational}
\end{equation}
All the branch cuts appear in the form of logarithms and
dilogarithms, and can be written as a sum over
a basis of scalar integrals --- bubbles $I_2^i$,
triangles $I_3^i$, and boxes $I_4^i$,
\begin{equation}
C_n\ =\ \sum_i d_i \, I_4^i\ +\ \sum_i c_i \, I_3^i\ +\ \sum_i b_i \, I_2^i \,.
\label{IntegralBasis}
\end{equation}
(Massive particles in the loop also require tadpole integrals.) 
We take all external momenta to be four dimensional, expressible
in terms of
spinors.  The coefficients of these
integrals, $b_i, c_i$, and $d_i$, as well as the rational remainder
$R_n$, are rational functions of spinor variables (in the form of
spinor products).  The \BlackHat{} library computes these coefficients
numerically, leveraging off recent analytic progress. In particular,
it exploits generalized unitarity~\cite{Z4Partons,BCFUnitarity}.
We use Forde's approach~\cite{Forde} to compute $b_i$ and $c_i$,
making use also of the subtraction approach to integral reduction
first introduced by Ossola, Papadopoulos and Pittau~\cite{OPP}.
To obtain the rational terms we
have implemented both loop-level on-shell recursion~\cite{Bootstrap},
and a ``massive continuation'' approach due to
Badger~\cite{Badger}, which is related to the $D$-dimensional generalized
unitarity~\cite{DdimUnitarity} approach of Giele, Kunszt and
Melnikov~\cite{GKM}. 

One-loop matrix element computations can suffer from numerical
instabilities. In \BlackHat, this problem is solved by detecting
pieces of the amplitude which do not have a sufficient accuracy and
recomputing them with higher precision using the multiprecision package
QD~\cite{QD}. This approach has the advantage of solving the problem
using the same approach for well-behaved points and for numerically
unstable ones. As discussed in
refs.~\cite{BlackHatI,W3jDistributions}, with a series of tests --- the
simplest of which checks whether the infrared divergences have the
proper values --- there is no need for {\em a priori\/} knowledge of what set
of circumstances can lead to instabilities. In each contribution
where precision loss is detected,
\BlackHat{} automatically switches to higher
precision, regardless of the underlying cause.  With on-shell methods
this happens infrequently and therefore has only a mild effect on the
overall computation time.

We have previously used these software tools to provide the first
phenomenologically useful NLO study of the production of a $W$ boson
in association with up to three jets~\cite{PRLW3BH,W3jDistributions}.
In this Contribution, we extend our previous studies with a more
detailed look at the question of scale choices; at aspects of the
polarization of $W$s produced at high $P_T$; and at a new distribution
displaying the probability of emitting a jet into a rapidity gap.  We
also present the first NLO results on \Zjjj-jet production at hadron
colliders, in a leading-color approximation designed to be accurate
within a few percent.  In all cases, we decay the vector boson to
leptons, $W^+\to l^+\nu_l$, $W^-\to l^-\bar\nu_l$, and $Z\to l^+l^-$,
using the appropriate vector boson linewidth.  We include the virtual
photon contribution to $l^+ l^-$ production.  Other recent
state-of-the-art NLO results may be found in
ref.~\cite{OtherRecentNLO}.  The production of \Wjjj\ jets has also
been computed at NLO using a leading-color approximation and
extrapolation~\cite{EMZW3j,MelnikovLHC}.

\section{Scale Choices}

The renormalization and factorization scales are not physical scales.
Physical quantities should be independent of them.  A dependence on
them is nonetheless present in theoretical predictions that are
truncated at a fixed order in perturbation theory.  At leading order, the
dependence arises solely through $\alpha_s$ and the
parton distributions, respectively.  We adopt the usual practice and
choose the two to be equal, $\mu_R = \mu_F = \mu$.  NLO results
greatly reduce the dependence compared to LO, but of course they do
not eliminate it completely.  We still need to choose this scale.  We
should expect a good choice for $\mu$ to be near a typical energy
scale for the observable we are computing, in order to minimize the
uncomputed logarithms in higher-order terms.  However, multi-jet
processes such as \Vjjx-jet production have many intrinsic scales, and
it is not clear {\em a priori\/} how to distill them into a single
number.  For any given point in the fully-differential cross section,
there is a range of scales one could plausibly choose.  For example,
one might choose the same fixed scale $\mu$ for all events.  However,
because there can be a large dynamic range in momentum scales
(particularly at the LHC, where jet transverse energies well above
$M_W$ are common), it is natural to pick the scale $\mu$
dynamically, event by event, as a function of the event's kinematics.

%%%%%%%%%%%%% FIGURE %%%%%%%%%%%%%%%%%%
\begin{figure}[tbh]
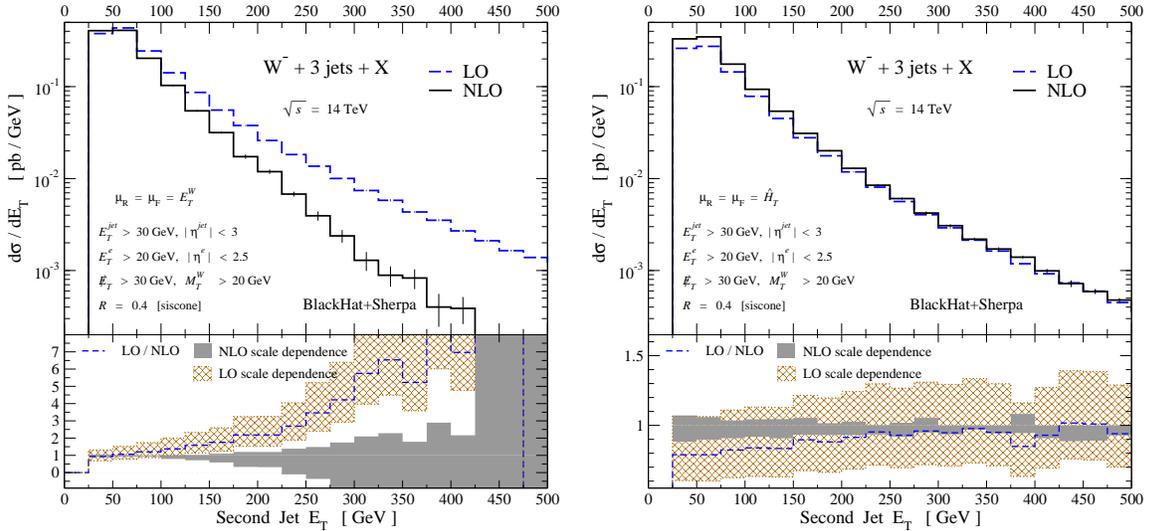

\begin{minipage}[b]{1.\linewidth}
\includegraphics[clip,scale=0.35]%
{plots/Wm3jLHC_ETWmu_siscone_eb_jets_jet_1_1_Et_2.eps}
\hskip0.31cm 
\includegraphics[clip,scale=0.35]%
{plots/Wm3jLHC_HTmu_siscone_eb_jets_jet_1_1_Et_2.eps}
\end{minipage}
\caption{LO and NLO predictions for the second jet $E_T$ distribution
in $W+3$ jet production at the LHC.  The only difference between the left
and right panels is the scale choice: $\mu = E_T^W$ on the left
and $\mu = \HTp$ on the right.  The former choice is clearly
problematic and should not be used in phenomenological studies.
The bottom panels show the LO and NLO predictions, varied by
a factor of two around the central scale, and divided by the NLO
value at the central scale.
}
\label{ScaleSecondJetETFigure}
\end{figure}
%%%%%%%%%%%%%%%%%%%%%%%%%%%%%

%%%%%%%%%%%%% FIGURE %%%%%%%%%%%%%%%%%%
\begin{figure}[tbh]
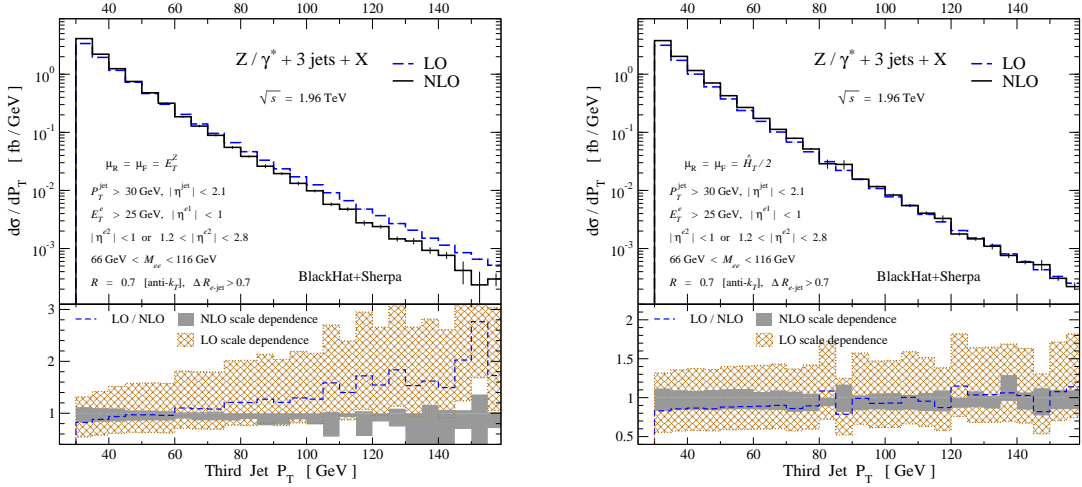

\begin{minipage}[b]{1.\linewidth}
\includegraphics[clip,scale=0.32]%
{plots/Z3j-Tev-ETZ_anti-kt-Pt30_jets_jet_1_1_pt_3.eps}
\hskip 1cm
\includegraphics[clip,scale=0.32]%
{plots/Z3j-Tev-HT_anti-kt-Pt30_jets_jet_1_1_pt_3.eps}
\end{minipage}
\caption{The NLO $P_T$ distribution of the third jet in \Zjjj-jet
production at the Tevatron. For the left panel the scale choice
$\mu = E_T^Z$ is used, and for the right panel $\mu = \HTp/2$.
Although the two NLO results are compatible, the LO results 
have large shape differences, illustrating that $\mu=\HTp/2$
is a better choice than $\mu = E_T^Z$
at the Tevatron as well.
The lepton and jet cuts match the CDF ones~\cite{ZCDF}.
} 
\label{ThirdJetTeVComparisonFigure}
\end{figure}
%%%%%%%%%%%%%%%%%%%%%%%%%%%%%

Previous studies (see {\it e.g.}~refs.~\cite{ZCDF,WCDF}) have used the
transverse energy of the vector boson, $E_T^V$, as the scale choice.
For many distributions at the Tevatron, this is satisfactory.
With the larger dynamic range at the LHC, the choice becomes
problematic. Indeed, for some observables, such as the
transverse-energy distribution of the second-hardest jet in \Wjjj-jet
production, shown in the left panel of \fig{ScaleSecondJetETFigure},
it goes disastrously wrong, leading to negative values of the
distribution for $E_T$ beyond 475 GeV.  Even at the Tevatron, the
scale choice $\mu = E_T^V$ is not necessarily a good one; for example,
with this choice, the left panel of \fig{ThirdJetTeVComparisonFigure}
displays a large change in shape between LO and NLO in the $P_T$
distribution of the third hardest jet in \Zjjj-jet production.  This
difficulty reflects the emergence of a large logarithm $\ln(\mu/E)$,
where $E$ is a typical energy scale, spoiling the validity of the
perturbative expansion.

%%%%%%%%%%%%% FIGURE %%%%%%%%%%%%%%%%%%
\begin{figure}[tbh]
\hskip 2.8 cm \includegraphics[clip,scale=0.55]{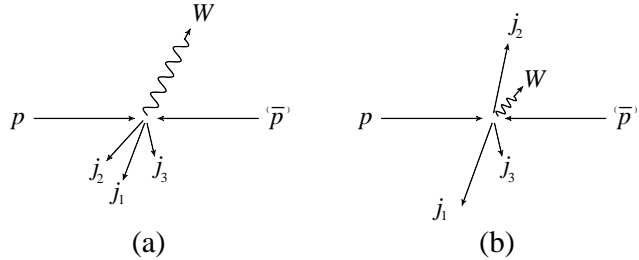}
\caption{Two distinct \Wjjj{} jet configurations with rather
different values for the $W$ transverse energy.
In configuration (a) an energetic $W$ balances the energy
of the jets, while in (b) the $W$ is relatively soft.
Configuration (b) generally dominates over (a) when the jet transverse
energies get large. }
\label{TwoConfigurationsFigure}
\end{figure}
%%%%%%%%%%%%%%%%%%%%%%%%%%%%%

To understand the problem with the scale choice $\mu = E_T^V$, consider the
two configurations depicted in \fig{TwoConfigurationsFigure}.
In configuration~(a), the $W$ has a transverse energy larger than
that of the jets, and accordingly sets the scale for the process.
In configuration~(b), the two leading jets roughly balance in $E_T$,
while the $W$ has much lower transverse energy.  Here, the $W$ scale
is too low, and not characteristic of the process.  In the tail
of the distribution, we expect configuration~(b) to dominate, because
it results in a larger second-jet $E_T$ for fixed center-of-mass
partonic energy; contributions from higher center-of-mass energies
will be suppressed by the fall-off of the parton distributions.

Can we choose a scale that treats the different final-state objects
more democratically?  The total partonic transverse energy,
\begin{equation}
\HTp\ =\ \sum_{{\rm partons\ }i} E_T^i\ +\ E_T^e\ +\ \ETslash\,,
\end{equation}
or a fixed fraction of it, is such a choice.  As we can see in the
right panels of
\figs{ScaleSecondJetETFigure}{ThirdJetTeVComparisonFigure}, this
choice results in stable and sensible NLO predictions --- and also in
a relatively flat ratio of the NLO and LO predictions.  For LO
predictions, it is better to use such a scale when NLO results are
unavailable.  A similar type of scale choice, based on the combined
invariant mass of the jets, has been motivated by soft-collinear effective
theory~\cite{Bauer}.  Local scales associated with ``branching histories''
as used in parton showers have recently been studied for \Wjjj-jet
production at LO~\cite{MelnikovLHC}.

\section{$Z$\,+\,Jets at the Tevatron}

At hadron colliders, $Z$ boson production manifests itself primarily
in either charged-lepton pair production, or the production of missing
transverse energy (when the $Z$ decays to neutrinos).  The latter
process is an important background to a wide variety of supersymmetry
searches (when no charged lepton is required), 
and to dark matter searches more generally.  The $l^+l^-$ mode has a
significantly lower rate, but it is an excellent calibration process,
as the $Z$ can be reconstructed precisely.  It is also an excellent process
for confronting NLO predictions with experimental data.

We have computed the NLO \Zjjja-jet production cross sections for 
the Tevatron ($p\pb$ collisions at $\sqrt{s}=1.96$~TeV), with the
$Z$ decaying into a charged lepton pair.
We applied the same cuts used by the CDF collaboration~\cite{ZCDF}
in their measurement of these processes for $Z\to e^+e^-$,
\begin{eqnarray}
&& P_T^\jet > 30~{\rm GeV}\,, \qquad
E_T^e > 25~{\rm GeV}\,,\qquad \Delta R_{e-\jet} > 0.7\,,
\qquad 66 < M_{e^+e^-} < 116~{\rm GeV} \,,
\nonumber\\
&& |\eta^\jet| < 2.1 \,, \qquad
|\eta^{e_1}| < 1\,,\qquad
|\eta^{e_2}| < 1 \quad{\rm or}\quad 1.2 < |\eta^{e_2}| < 2.8\,,
\label{ZJetCuts}
\end{eqnarray}
where the electron cuts apply to both electrons and positrons,
and the jet cuts apply to all jets.  We cut on the jet
pseudo-rapidity $\eta$ rather than CDF's cut on rapidity $y$;
the two cuts coincide at LO but differ slightly at NLO.
We employed three different
infrared-safe jet algorithms~\cite{Jetography}, SISCone
(with merging parameter $f = 0.75$), $k_T$ and anti-$k_T$,
all with $R=0.7$.  Production of an $l^+l^-$ pair can also be 
mediated by a virtual photon; we include these contributions as well,
although they are suppressed by the cut on the lepton-pair
invariant mass $M_{e^+e^-}$.

%%%%%%%%%%%%% FIGURE %%%%%%%%%%%%%%%%%%
\begin{figure}[tbh]
\begin{center}
\includegraphics[clip,scale=0.4]%
{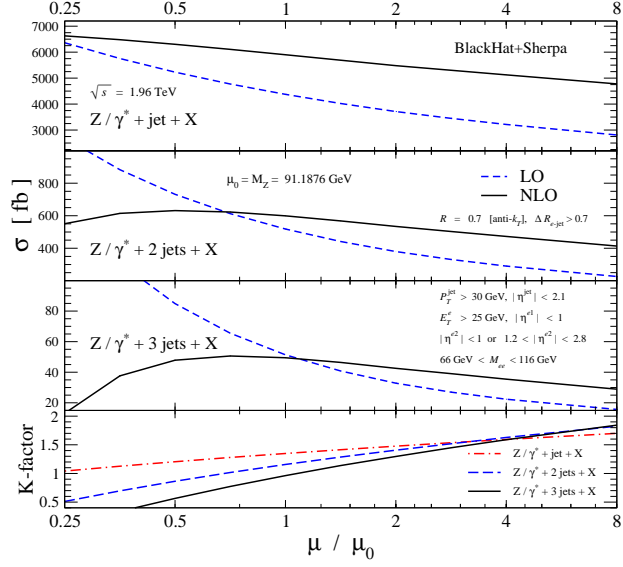}
\end{center}
\vskip -.3 cm 
\caption{The scale dependence of the cross section for \Zjjja-jet
production at the Tevatron, for the anti-$k_T$ jet algorithm
using a leading-color approximation with $n_f$ terms,
as a function of the common
renormalization and factorization scale $\mu$, with $\mu_0 = M_Z$.
The bottom panel shows the $K$ factors, or ratios between NLO 
and LO results, for the three cases.}
\label{ZJetsScaleVariationFigure}
\end{figure}
%%%%%%%%%%%%%%%%%%%%%%%%%%%%%

\Fig{ZJetsScaleVariationFigure} shows how the \Zjjja-jet 
cross section depends on a fixed scale $\mu$, independent of
the event kinematics, for the anti-$k_T$ algorithm and with 
the cuts~(\ref{ZJetCuts}).
Here choosing $\mu\approx M_Z$ is appropriate, 
because the cross section is dominated by low-$P_T$ jets.
The upper three panels show the
scale dependence of the cross section
at NLO, compared to that at LO, in
\Zj-, {\Zjj-,} and \Zjjj-jet production, respectively.  They illustrate
the lessened dependence at NLO.
The bottom panel
shows the ratio of NLO to LO results for all three cases, 
demonstrating
the increasing sensitivity to scale variations
at LO with increasing number of
jets.  This is expected, because there is an additional power of
$\alpha_s(\mu)$ multiplying the LO cross section for each additional jet.
Accordingly, the impact of an NLO calculation also grows with
the number of jets.  The results for the $k_T$
and SISCone algorithms (not shown) are similar.

%%%%%%%%%%%%% FIGURE %%%%%%%%%%%%%%%%%%
\begin{figure}[tbh]
\hskip 3 cm 
\includegraphics[clip,scale=0.34]%
{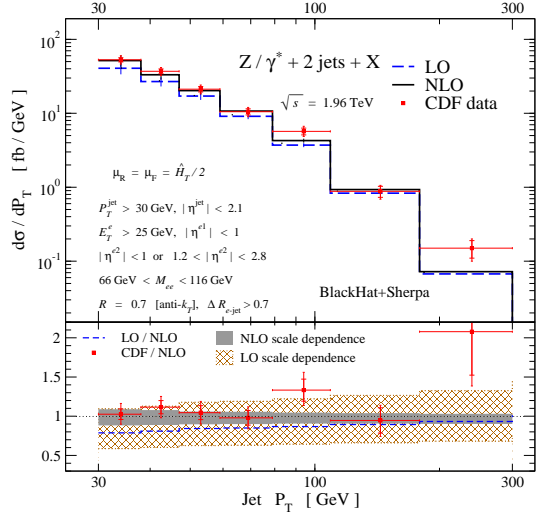}
\caption{The second-jet $P_T$ distribution for $Z+2$ jets at LO and NLO
compared against CDF data~\cite{ZCDF}. }
\label{ZPlus2JetFigure}
\end{figure}
%%%%%%%%%%%%%%%%%%%%%%%%%%%%%

\Fig{ZPlus2JetFigure} compares the theoretical
predictions for the second-jet $P_T$ distribution in \Zjj-jet
production with data from CDF~\cite{ZCDF}.  CDF
used the midpoint algorithm~\cite{MidPoint}.
This algorithm is infrared unsafe for \Zjjj-jets at NLO, so we 
use
infrared-safe ones instead.  \Fig{ZPlus2JetFigure} shows results for the
anti-$k_T$ algorithm; the other two algorithms yield similar results.
It is worth noting that CDF did not attempt to ``deconvolve'' the
hadronization corrections~(estimated using Pythia) from their measured
data; rather, they provided a table of hadronization corrections.
This is helpful because it will allow for future improvements to
hadronization models to be taken into account in theoretical
predictions.  Accordingly, we have used these hadronization
corrections to generate a complete prediction from the LO and NLO
perturbative predictions. The hadronization corrections are significant
for low $P_T$, on the order of 20\,\% at 30 GeV,
and become rather small at larger jet transverse
momenta.  As expected, the LO scale-dependence band is much larger
than the NLO one.  Excepting perhaps the last bin, the agreement between
the NLO prediction and the data is quite good, especially
given the different jet algorithms.

\begin{figure}[tbh]
\begin{center}
\includegraphics[clip,scale=0.46]%
{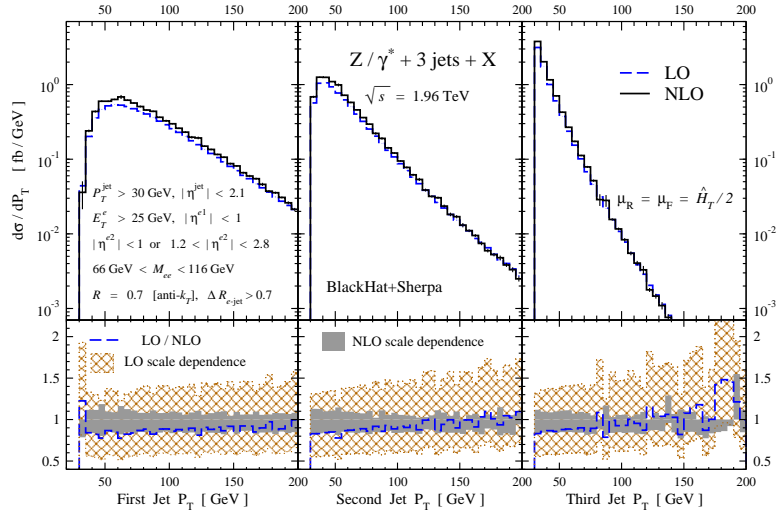}
\end{center}
\vskip -.3 cm 
\caption{The LO and NLO $P_T$ distributions for \Zjjj-jet production
for the leading, second and third jet, for the anti-$k_T$ algorithm
and scale choice $\mu = \HTp/2$. The thin vertical bars in the top
panels indicate the integration errors.}
\label{ZPlus3Figure}
\end{figure}

\Fig{ZPlus3Figure} gives our predictions for the three jet $P_T$
distributions in \Zjjj-jet production, using the anti-$k_T$ jet
algorithm.  With the choice of scale $\mu = \HTp/2$, only minor shape
changes are visible between LO and NLO, for all three distributions.
The NLO plots are based on a leading-color approximation along the
lines of refs.~\cite{PRLW3BH,W3jDistributions}, except
that pieces proportional to the number of light quark flavors ($n_f$)
are included.  We expect this approximation to be valid to a few
percent.

%%%%%%%%%%%%% FIGURE %%%%%%%%%%%%%%%%%%
%%%%%%%%%%%%%%%%%%%%%%%%%%%%%

\section{$W$ Polarization at the LHC}

%%%%%%%%%%%%% FIGURE %%%%%%%%%%%%%%%%%%
\begin{figure*}[tbh]
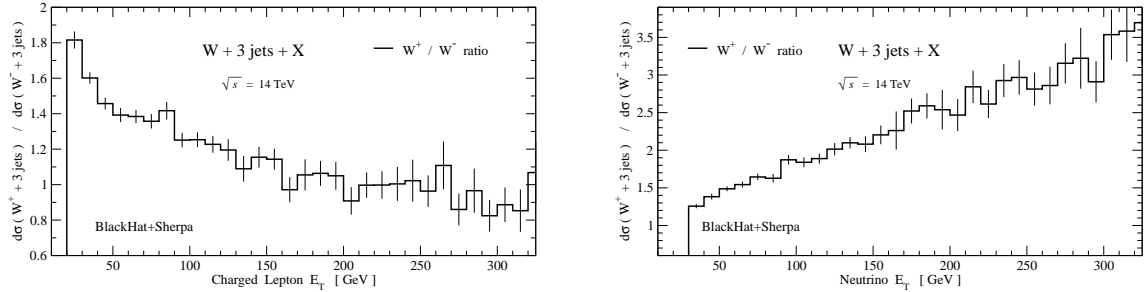

\begin{minipage}[b]{1.\linewidth}
\includegraphics[clip,scale=.3]%
{plots/Wp3jLHCHT_over_Wm3jLHCHT_central_mu_siscone_PTchargedlepton.eps}
\hfill
\includegraphics[clip,scale=.3]%
{plots/Wp3jLHCHT_over_Wm3jLHCHT_central_mu_siscone_PTneutrino.eps}
\end{minipage}
\caption{The left panel shows the ratio of the charged-lepton $E_T$
distributions at the LHC for $W^+$ and $W^-$ production in
association with at least three jets, computed at NLO. 
The right panel shows the corresponding ratio for the
neutrino $E_T$, or equivalently $\ETslash$.}
\label{WPTchargedneutralleptoni_ratio_Figure}
\end{figure*}
%%%%%%%%%%%%%%%%%%%%%%%%%%%%%%%%%%%%%%%

As noted in ref.~\cite{W3jDistributions}, at the LHC the $E_T$
distributions of the daughter leptons show a surprisingly strong shape 
dependence on whether they come from a $W^+$ or a $W^-$, independent
of the number of jets.
\Fig{WPTchargedneutralleptoni_ratio_Figure} shows the ratio of the NLO
transverse energy distributions for the $W^\pm$ boson decay products
in inclusive \Wjjj-jet production at the LHC,
charged leptons in the left panel and neutrinos in the right
panel. The differences between $W^+$ and $W^-$
distributions are quite dramatic.  The left panel
shows a large ratio for $W^+$ to $W^-$ at small $E_T^e$ which declines
at larger $E_T^e$.  In contrast, the corresponding ratio for the
$E_T^\nu$, or equivalently the missing transverse energy
$\ETslash$ in the event, starts somewhat smaller but
increases rapidly with $E_T$.  The significant difference in behavior
between $W^+$ and $W^-$ suggests a means for separating $W$
bosons produced in top quark decays from those produced from light quarks;
the $W$s from top decays do not exhibit a similar phenomenon.

%%%%%%%%%%%%% FIGURE %%%%%%%%%%%%%%%%%%
\begin{figure}[tbh]
\begin{minipage}[b]{1.\linewidth}
\hskip -.05 cm
\includegraphics[clip,scale=0.4, angle=270]%
{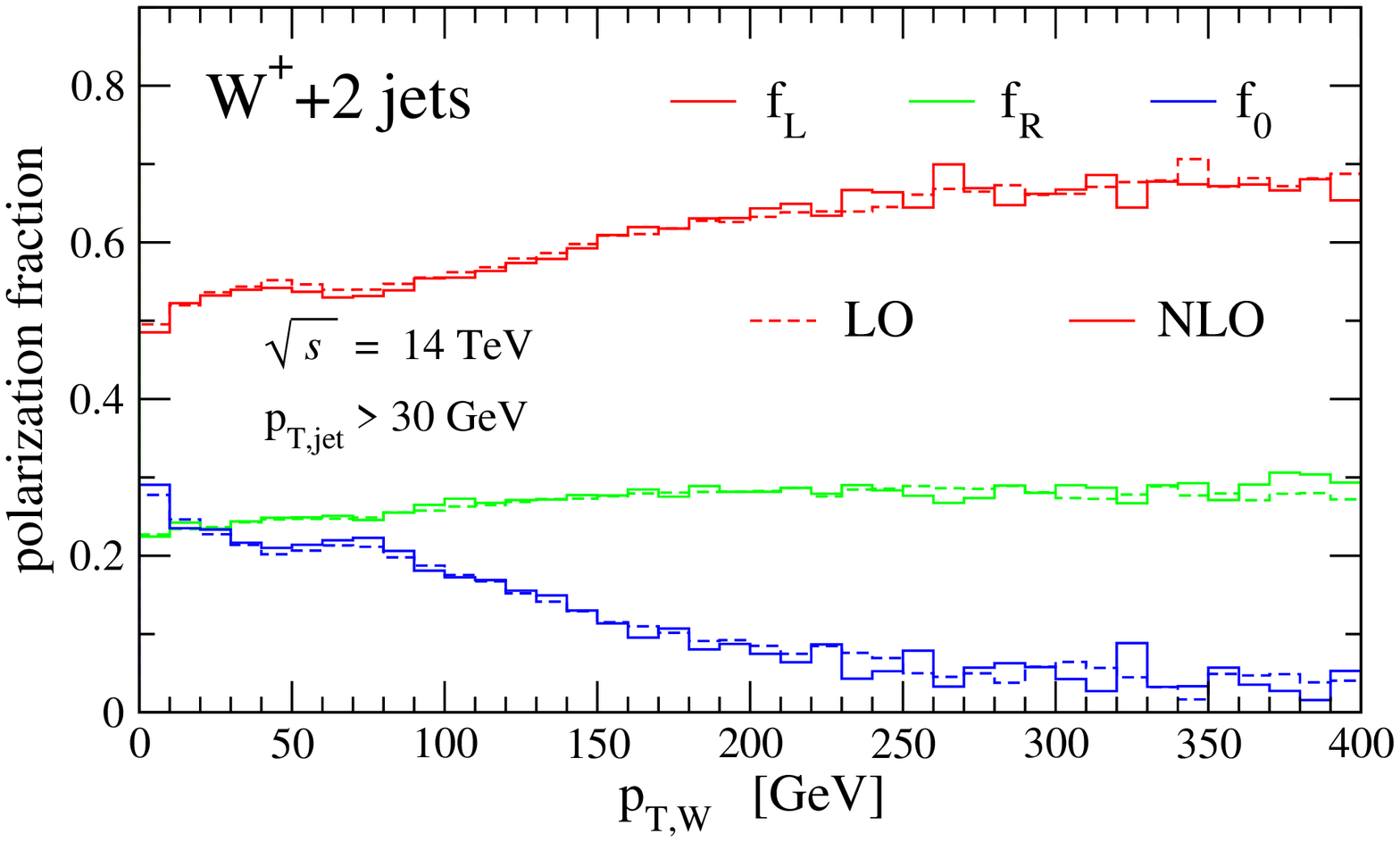}
\hskip .18 cm
\includegraphics[clip,scale=0.4, angle=270]%
{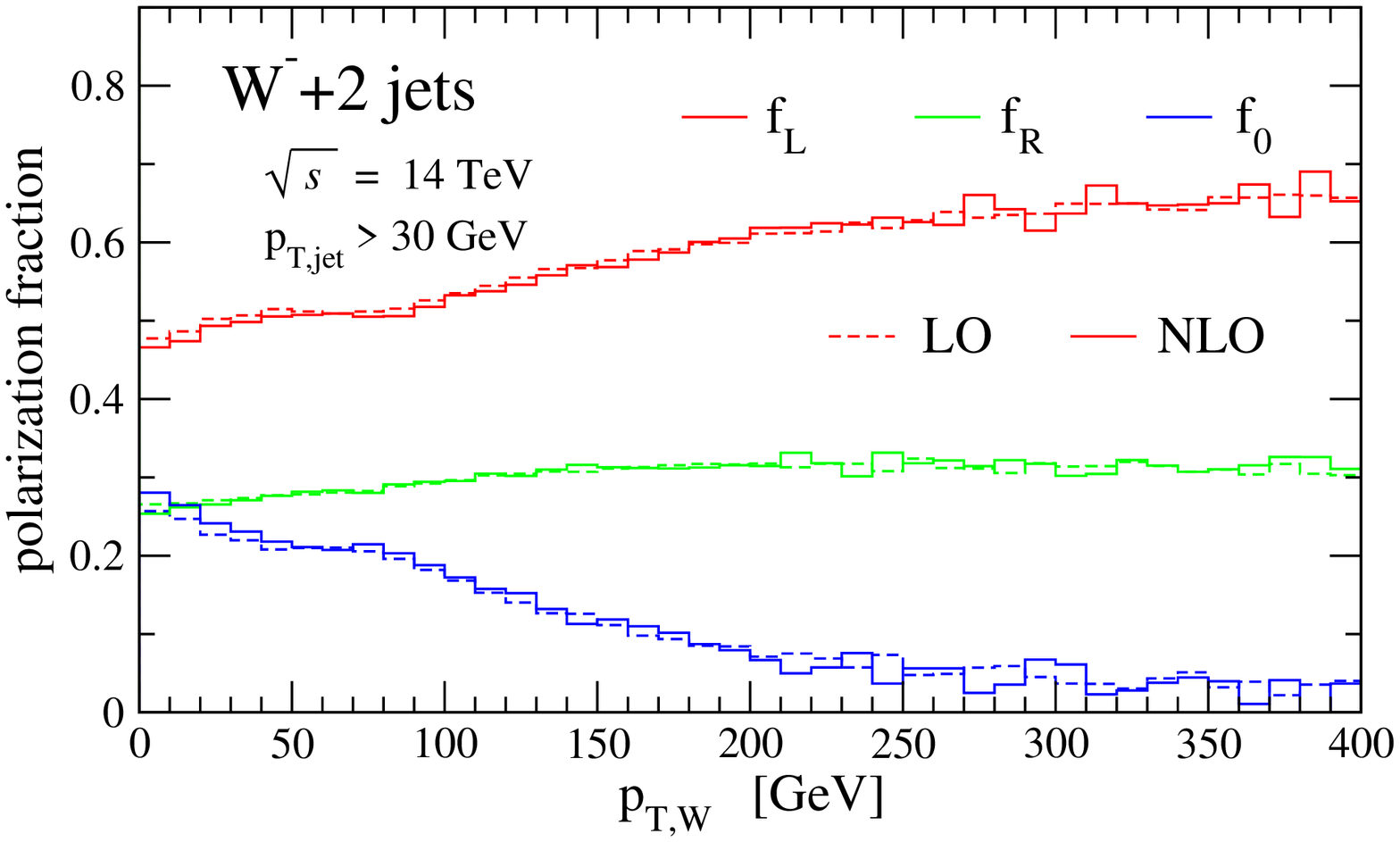}
\end{minipage}
\caption{The LO and NLO predictions for polarization fractions of the
left-handed, $f_L$ (top curve), right-handed $f_R$ (middle curve) and
longitudinal $f_0$ (bottom curve) fractions for \Wjj{} jets at the LHC.
The left panel gives the polarization for $W^+$ and the right panel
for $W^-$.  For high transverse momentum, $P_{T,W}$, the $W$ bosons
become predominantly left-handed.
 }
\label{WPolarization}
\end{figure}
%%%%%%%%%%%%%%%%%%%%%%%%%%%%%

This disparate behavior is explained by a net left-handed polarization
for both $W^+$ and $W^-$ at high transverse momentum.  This effect is
easily visible at LO, and it does not get washed out at NLO.  In
\fig{WPolarization}, we give the fraction of $W$ bosons in each of the
three polarization states, left-handed, right-handed and longitudinal
($f_L, f_R, f_0$, respectively) for \Wjj-jet production
at the LHC, at both LO and NLO. 
As seen in the figure, at
high transverse momentum the $W^\pm$ bosons are preferentially left
handed. Although the cross-sections for $W^+$ and $W^-$ are rather
different, their polarizations are nearly identical.
Interestingly, we also find that when the $W$s
have a transverse momentum of more than 50 GeV, the polarization is
quite independent of the jet transverse energy cuts.  With $W^\pm$
bosons left-hand polarized at large $E_T^W$, the $W^+$
tends to emit the left-handed neutrino forward relative to its
direction of motion (resulting in a larger transverse energy) and the
right-handed positron backward (smaller transverse energy).  
In contrast,  the $W^-$ prefers to emit the left-handed electron forward.
At high $E_T$, such decays produce an
enhancement in the neutrino $E_T$ distribution and a depletion in the
charged-lepton distribution, for $W^+$ relative to $W^-$, consistent
with the results displayed
in~\fig{WPTchargedneutralleptoni_ratio_Figure}.  We note that this
phenomenon is distinct from the well-known dilution of the $W$
rapidity asymmetry at the Tevatron, when passing to the decay
lepton, which can be explained using angular momentum
conservation solely along the beam axis~\cite{ESW}.

\section{Emission into Rapidity Gaps}

In previous work~\cite{WCDF}, we provided the first NLO study of the
probability of emitting a third jet in \Wjj-jet events, 
as a function of the rapidity interval between two leading-$E_T$ jets
at the LHC. This distribution was studied earlier at LO at the Tevatron
and compared to CDF data~\cite{CHS}.
Jet emission probabilities are relevant to Higgs
searches in vector-boson fusion~\cite{VBF}, in which color-singlet
exchange leads to a paucity of jet radiation in the central region
between two forward tag jets.  On the other hand,
QCD backgrounds with color exchange, as in \Wjj-jet production,
will generally lead to significant jet radiation.

%%%%%%%%%%%%% FIGURE %%%%%%%%%%%%%%%%%%
\begin{figure}[tbh]
\begin{center}
\includegraphics[clip,scale=0.34]%
{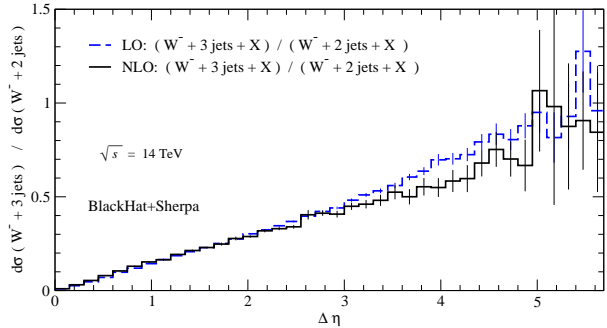}
\end{center}
\vskip -.3 cm
\caption{The ratio of the inclusive \Wmjjj-jet cross section 
to the \Wmjj-jet cross section as a function
of the pseudorapidity separation $\Delta \eta$ between the two
most widely separated jets that pass the cuts.  The solid (black) 
line gives the NLO result, while the dashed (blue) line gives
the LO results.  
  }
\label{EtaEmissionDistributionFigure}
\end{figure}
%%%%%%%%%%%%%%%%%%%%%%%%%%%%%

To mimic vector-boson fusion searches, however,
the appropriate tag jets are not the two hardest ones (by $E_T$), 
but rather the two most separated in pseudorapidity.
Therefore, in \fig{EtaEmissionDistributionFigure} we
present the ratio of the \Wmjjj-jet cross section to the \Wmjj-jet
cross section as a function of the pseudorapidity separation 
$\Delta\eta$ between the two most separated jets.  The emission probability
rises roughly linearly with $\Delta\eta$. 
The NLO result is somewhat less than the LO one at large $\Delta \eta$.
(The ratio for $W^+$ is quite similar.)  This plot is
similar to one for Higgs production in association with
jets~\cite{AndersenHiggs}, obtained from high-energy factorization
considerations.  It would be interesting to compare results obtained
in this way to NLO results for the same quantities.

\section{Conclusions}

In this Contribution we presented some new results for \Wjjj-jet
production obtained from \BlackHat{} combined with SHERPA, 
expanding on earlier 
scale-dependence studies~\cite{PRLW3BH,W3jDistributions}.  We also
demonstrated that $W$ bosons produced at large $P_T$ are
indeed polarized left-handed, explaining an asymmetry between $W^+$ and $W^-$
in the transverse energy distributions of the daughter leptons. Because
$W$s from top decays do not exhibit this polarization effect, it may
prove effective for distinguishing such $W$s from ones
produced by light quarks.  We presented the
first NLO study of the probability of emitting a third jet between the
two most widely separated jets in \Wjj-jet production.
We also presented the first NLO results for \Zjjj-jet production.
We observed that even at the Tevatron, choosing the renormalization
and factorization scale to equal the vector boson transverse energy 
is not a particularly good choice, as it induces large shape changes
between LO and NLO.

A publicly available version of \BlackHat{} is in preparation and is
currently being tested in diverse projects
(see {\it e.g.}~ref.~\cite{RikkertContribution}).
This version uses the proposed
Les Houches interface for one-loop matrix elements. It has been tested
with both C++ and Fortran clients. The public version will provide all
processes that have been carefully tested with the full \BlackHat{} code.

In the more distant future, the next benchmark process for 
\BlackHat{}\,+\,SHERPA is the production 
of a $W$ boson in association with four jets at NLO.
Using the techniques described above, the virtual part of the NLO
cross section seems within reach.  Computing the real
emission matrix elements, and integrating them over the seven-particle
phase space (including the decay of the vector boson) appears to be
rather challenging with the current tools, due to the large number of
integration channels.  It is interesting to note that in this case
the bottleneck no longer seems to be the virtual contributions to the
cross section.

The results summarized here are indicative of the type of physics
that can be carried out using \BlackHat{} in conjunction with SHERPA.
We look forward to comparing predictions from these tools to the
forthcoming LHC data.

\section*{Acknowledgments}

\vskip -.3 cm 
We thank Jeppe Andersen, Rikkert Frederix, and Markus Stoye
 for helpful conversations.
This research was supported by the US Department of Energy under contracts
DE--FG03--91ER40662, DE--AC02--76SF00515 and DE--FC02--94ER40818.
DAK's research is supported by the European Research Council under
Advanced Investigator Grant ERC--AdG--228301.  This research used
resources of Academic Technology Services at UCLA, PhenoGrid using the
GridPP infrastructure, and the National Energy Research Scientific
Computing Center, which is supported by the Office of Science of the
U.S. Department of Energy under Contract No. DE--AC02--05CH11231.

\end{document}